\documentstyle[12pt,epsf]{article} 
\newcommand{\be}{\begin{equation}} 
\newcommand{\en}{\end{equation}}
\newcommand{\bea}{\begin{eqnarray}}
\newcommand{\ena}{\end{eqnarray}}

\newcommand{\hbo}{\hbox to 1 true cm {\hfill } }

\def\dslash{\partial\kern-.5em\slash}
\def\kslash{k\kern-.5em\slash}
\def\pslash{p\kern-.5em\slash}

\begin{document} 
\vglue 1truecm
  
\vbox{ UNITU-THEP-11/1996 
\hfill July 24, 1996 
}
  
\bigskip 
\centerline{\large\bf Quark Confinement due to Random 
   Interactions\footnote{Supported by DFG under contracts La 932/1-2. } } 
  
\bigskip
\centerline{ K.\ Langfeld } 
\bigskip
\centerline{ Service de Physique Th\'eorique, C.E. Saclay, }
\centerline{ F-91191 Gif-sur-Yvette Cedex, France. } 
\centerline{and} 
\centerline{ Institut f\"ur Theoretische Physik, Universit\"at 
   T\"ubingen\footnote{present address; email: {\tt 
   langfeld@ptsun1.tphys.physik.uni-tuebungen.de}}
 }
\centerline{D--72076 T\"ubingen, Germany.}
\bigskip

\begin{abstract}
\begin{center} 
\parbox{5in}{ \small 
On the level of an effective quark theory, we define confinement 
by the absence of quark anti-quark thresholds in correlation functions. 
We then propose a confining Nambu-Jona-Lasinio-type model. 
Its four fermion interaction in the color channel is mediated by a 
random color matrix. 
We study the model's phase structure as well as its behavior 
under extreme conditions, i.e.\ high temperature and/or high density. 
In order to study the interplay of the proposed mechanism of confinement 
and asymptotic freedom, we consider an extended Gross-Neveu model. 
Quark anti-quark thresholds are proven to be absent in the scalar correlation 
function implying that the decay of the scalar meson into free quarks is 
avoided. 
The transition from the low energy to the high energy perturbative regime 
is found to be smooth. Perturbation theory provides a good description 
of Green's functions at high energies, although there is no 
quark liberation. 
} 
\end{center}

\end{abstract}

\vskip 1.5cm
{\bf 1. Introduction } 

Since the invention of QCD as the theory of strong interactions in 
the early seventies, we observe a steady increase in knowledge of its low 
energy properties. Instantons trigger the spontaneous breakdown of 
chiral symmetry~\cite{ca78} and give rise to the $\eta $-$\eta '$ 
mass splitting~\cite{wi79}. Quark confinement is observed in lattice 
QCD, since Wilson's area law is satisfied~\cite{eng87}. Even a qualitative 
understanding of quark confinement is available from the dual 
super-conductor picture~\cite{bak91}. In the so-called Abelian gauges 
Yang-Mills theory possesses monopoles. If these monopoles condense, 
a dual Meissner effect takes place expelling electric flux from 
the vacuum. A flux tube is formed between static color sources giving 
rise to a linearly rising confinement potential. A recent 
success~\cite{sei94} of Seilberg and Witten supports these ideas. 
They showed that the monopoles of $N=2$ (non-confining) super-symmetric 
Yang-Mills theory start to condense, if the theory is explicitly 
broken down to $N=1$ (confining) SUSY. 

Despite this success in understanding low energy QCD, it is still not 
possible to describe hadron properties from first principles. One 
must resort to non-linear meson theories or effective quark models, which 
are inspired by the QCD symmetry behavior. The model of 
Nambu-Jona-Lasinio~\cite{nambu} (NJL) is one of the most economical models 
that possess the essence of dynamical symmetry breaking that is the 
hall-mark of modern field theories and has enjoyed an impressive 
phenomenological success in hadron physics~\cite{appli}. Unfortunately, 
the NJL-model does not include quark-confinement. This implies that, 
in particular, a meson can decay into a free quark anti-quark pair, if 
the decay is allowed by kinematics. This un-physical decay severely limits 
the predictive power of the NJL-model~\cite{ja92}. 

Here, I will report results~\cite{la96a,la96b} which Mannque Rho and I 
have obtained from an extended 
NJL-model which incorporates confinement. Quark confinement will be thereby 
defined by the absence of quark thresholds in correlation functions. 
The model's phase structure will be studied as well as its behavior under 
extreme conditions, i.e. high temperature and/or density. 
Finally, I will investigate the model in two dimensions, which then is an 
extended version of the Gross-Neveu model. This model is renormalizable and 
asymptotically free. It explains why perturbation theory is good at high 
energies, although there are no free quarks.

\medskip 
{\bf 2. Random interactions and quark confinement } 

If electrons are scattered off by randomly distributed impurities, 
the electron wave function gets localized~\cite{anderson}. One might 
question whether a similar effect leads to a confinement of quarks. Is 
there are a random interaction which stems form the gluonic interaction? 
The answer to this question is not known. There is, however, a first 
hint from a reformulation of QCD in terms of field strength~\cite{sch90}. 
In the strong coupling limit, a quark current-current interaction 
occurs which is mediated by the field strength of the gluonic 
background field~\cite{sch90,la91}. Since all orientations of the 
background field contribute with equal parts, an average over 
all orientations is understood. This gives rise to a random 
interaction in configuration space\footnote{The kind of random interaction 
is, however, conceptually different from that in solid state physics.}. 

We do not want to derive a low energy effective quark theory from 
first principles, but we will use the field strength approach to 
Yang-Mills theory to motivate an extended version of the NJL-model. 
We write the generating functional for mesonic Green's functions 
in Euclidean space as 
\bea 
W[\phi ] &=& \left\langle \ln \int {\cal D} q \; {\cal D } \bar{q} \; 
\exp \left\{ - \int d^{4}x \; [ 
{\cal L } \, - \, \bar{q}(x) \phi (x) q(x)] \; \right\} 
\right\rangle _{ O } \; , 
\label{eq:1} \\ 
{\cal L }  &=& \bar{q}(x) ( i \dslash + im ) q(x) \; + \; \frac{G_0}{2} 
[ \bar{q} q(x) \, \bar{q} q(x) \, - \, \bar{q} \gamma _5 q(x) \, 
\bar{q} \gamma _5 q(x) ] 
\label{eq:2} \\ 
&+& \frac{1}{2} [ \bar{q} \tau ^{\alpha } q(x)  \, G^{\alpha \beta } 
\bar{q} \tau ^{\beta } q(x) \, - \, 
\bar{q} \gamma _5 \tau ^{\alpha } q(x)  \, G^{\alpha \beta } 
\bar{q} \gamma _5 \tau ^{\beta } q(x)  ] \; , 
\nonumber 
\ena 
where $m$ is the current quark mass. We assume that
the quark interaction is given by a color-singlet four-fermion 
interaction of strength $G_0$ and a color-triplet interaction 
mediated by a positive definite matrix $G^{\alpha \beta }$, 
which represents gluonic background fields. An average 
of all orientations $O$ of the background field $G^{\alpha \beta }$, 
transforming as $G'=O^{T} G O$ with $O$ being a $3 \times 3$ 
orthogonal matrix, is understood in (\ref{eq:1}) to 
restore global $SU(2)$ color symmetry.  

The model (\ref{eq:1}) should be regarded as an effective low 
energy quark theory which is truncated at the level of energy dimension 
six operators and which is therefore valid up to a momentum cutoff 
$\Lambda $. 

In order to handle the random interaction in (\ref{eq:1}), we 
calculate the Green's function of interest for a fixed background 
interaction $G^{\alpha \beta }$ and average the result over all 
orientations $O$. For this purpose, the quark propagator $S(k)$ for 
fixed strength $G^{\alpha \beta }$ is necessary. Standard 
methods~\cite{nambu} yield 
\be 
S(k) \; = \; \frac{ 1 }{ \kslash \, + \, i(M_0 + i 
M^{\alpha } \tau ^{\alpha } ) } \; , 
\label{eq:e5} 
\en 
where the constituent quark mass in the color triplet channel, i.e. 
i$M^a$, is reminiscent of the background field $G^{\alpha \beta }$. 
The particular value for $M_0$ and $M^a$ is obtained as function 
of the interaction strengths from the gap-equations~\cite{la96a,la96b}. 
It turns out that there is always a solution with $M^a=0$. In this case, 
the model is reduced to the NJL-model (NJL-phase). The most striking 
feature is that there are in addition solutions where the mass in 
the color triplet channel is purely imaginary (i.e. $M^a$ real). This 
phase has lower action density and therefore constitutes the vacuum. 

\begin{figure}[t]
\centerline{ 
\hspace{2cm}
\epsfxsize=6.5cm
\epsffile{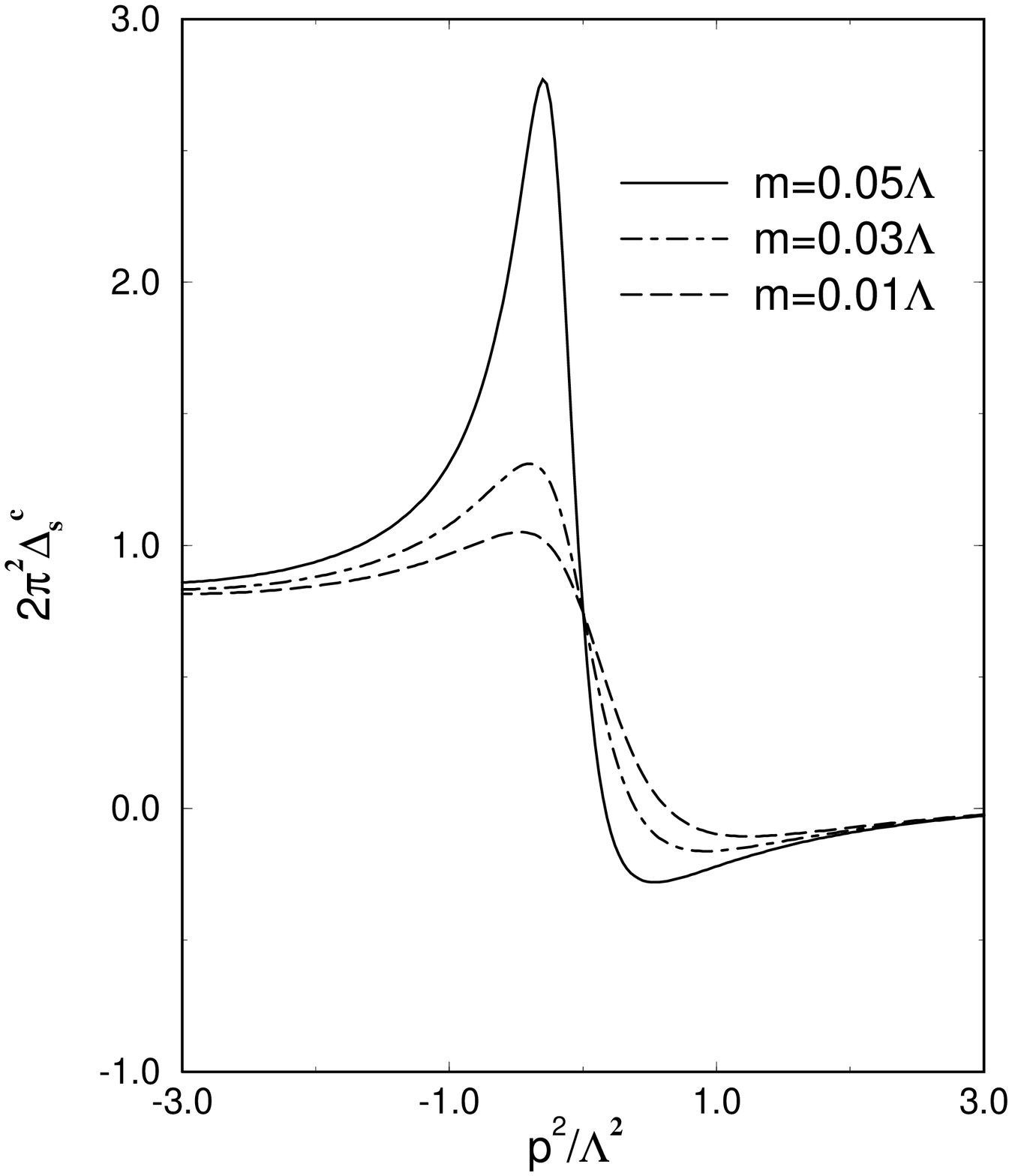} 
\hspace{2cm}
\epsfxsize=8cm
\epsffile{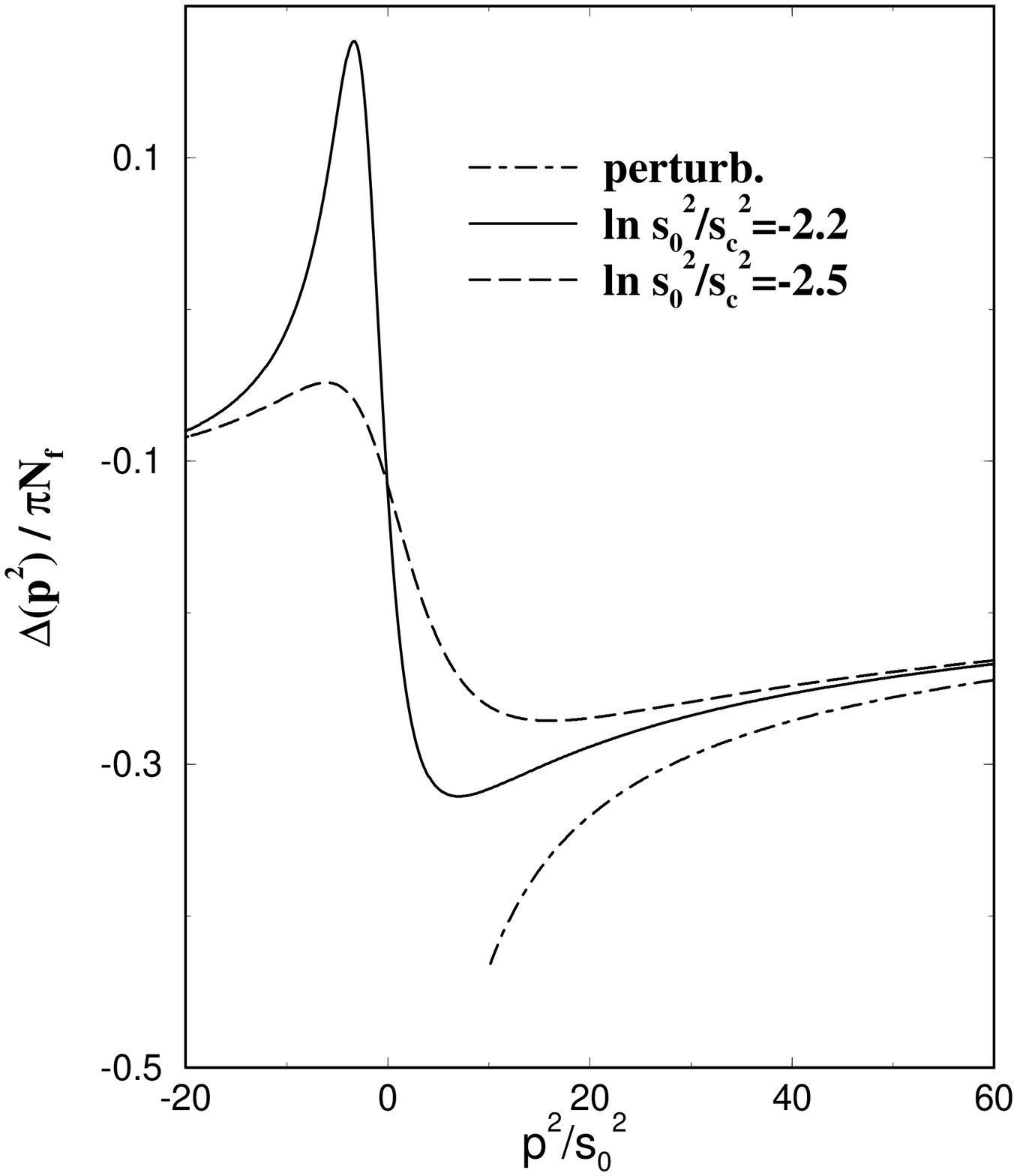} 
}
\vspace{.5cm} 
\caption{ The scalar correlation function as function of the 
   Euclidean momentum transfer for the extended NJL-model (left) and 
   the extended Gross-Neveu model (right). } 
\label{fig:1} 
\end{figure} 
Before we will discuss the phase structure of the model, we briefly 
investigate the physical consequence of the complex mass. 
For this purpose, we study the scalar correlation function in 
Euclidean space, i.e. 
\be 
\Delta _s (p) \; = \; 
\int d^{4}x \; e^{-ipx} \; \left\langle \bar{q}q(x) \; \bar{q}q (0) 
\right\rangle \; . 
\label{eq:9} 
\en 
Let us first consider the NJL-phase ($M^2:=M^aM^a=0$). The scalar 
correlation function exhibits a pole at $p^2= -4 M_0$, which 
signals the existence of a scalar meson. For $p^2< - 4 M_0^2$ the 
correlator (\ref{eq:9}) develops an imaginary part, which describes 
the decay of the scalar meson into free quarks. It is this unphysical 
threshold which constrains the applicability of the standard NJL-model. 

The situation is completely different in the phase with $M^2\not=0$. 
The scalar correlator $\Delta _s (p) $ is shown in figure \ref{fig:1}. 
No pole occurs. The scalar channel is closed. This result does 
not contradict experiments since no clear signal of a scalar meson 
is observed in the spectrum. The most important observation is, however, 
that the imaginary parts cancel. No unphysical thresholds 
occur~\cite{la96a,la96b}. In the following, 
we will refer to this phase as confining phase. The order parameter 
which distinguishes this phase from the NJL-phase is $M^2$. 
It turned out that the model is driven towards the NJL-phase, if the 
current mass is increased at fix interaction strengths. Figure \ref{fig:1} 
shows that the peak structure becomes more sharp in this case. 
In fact, the peak turns into a pole and an imaginary part develops, if the 
cross-over to the NJL-phase is reached. 

Finally, I summarize the results~\cite{la96a,la96b} at finite temperature 
and finite density, respectively. Let us choose the parameters that the 
confining phase is realized at zero temperature. Increasing the temperature 
(density) decreases the constituent mass in the color singlet channel 
$M_0$ and enhances $M^2$. At a critical temperature (density), the 
confinement order parameter $M^2$ suddenly drops to zero and a 
deconfinement phase transition takes place. This phase transition 
is accompanied by a restoration of chiral symmetry (for a wide parameter 
range). One observes that the critical temperature (density) is 
nearly insensitive to a change in the current mass. This implies that 
the critical temperature (density) in this model is almost the same for 
all quark flavors.

\medskip 
{\bf 3. Why is perturbation theory good, although there are no free 
quarks? } 

In order to address this question, we implement the above 
confinement mechanism in an extended version of the Gross-Neveu 
model, which is essentially the model (\ref{eq:1}) in two space-time 
dimensions. To be specific, the Euclidean generating functional is 
\bea 
Z[j] &=& 
\left\langle \ln \int {\cal D} q \; {\cal D} \bar{q} \; \exp \left\{ 
- \int d^{2}x \; [ L \; - \; j(x) \, g \bar{q}(x) q(x)  ] \, \right\} 
\right\rangle _{O} \; , 
\label{eq:g1} \\ 
L &=& \bar{q}(x) \left( i\dslash + im \right) q(x) \, + \, 
\frac{ g }{2 N_f } (\bar{q}(x) q(x) )^{2} \, + \, 
\frac{1}{2 N_f} \bar{q} \tau ^{a} q(x) \, \bar{q} \tau ^b q(x) \, 
G^{ab} \; , 
\label{eq:g2} 
\ena 
where $m$ is the current quark mass. 
The quarks fields $q(x)$ transform under a global $O(N_f)$ flavor 
and a global SU(2) (color) symmetry. $g$ and $G^{ab}$ are now dimensionless 
coupling constants in the color singlet and the color triplet channel 
respectively. As in the previous case, an average over all orientations 
$O$ of the background field $G^{ab}$ is understood. 
The crucial point is that this model is renormalizable and, in addition, 
asymptotically free. It is therefore suitable to study the interplay 
of asymptotic freedom and the non-liberation of quarks due to 
confinement. 

The model can be exactly solved in the large $N_f$ limit (for details 
see~\cite{la96b}). Due to dimensional transmutation, the dimensionless 
parameter $g$ is turned into a scale $s_0$, and a second scale $s_c$ 
occurs which characterizes the interaction strength in the color 
triplet channel. If we measure all physical quantities in units of 
the renormalization group invariant $s_0$, the only parameter of the 
model is $\ln s_0^2/ s_c^2$. 

It turns out, that the phase structure of the model is quite similar 
to that of the 4-dimensional model (\ref{eq:1}). In particular, 
the quark propagator for a fixed background field has again the 
structure (\ref{eq:e5}). If the interaction 
strength in the color triplet channels exceeds a critical value, i.e. 
$s_c > 2.71828 \ldots \, s_0 $, a confining phase with $M^2 \not= 0$ 
occurs. The scalar correlation function of the present model is 
also shown in figure \ref{fig:1}. No pole and no imaginary parts 
occur. The new feature is that we are now able to compare the full 
(large $N_f$) result with that of perturbation theory. One observes a 
good agreement at large $p^2$ as expected from asymptotic freedom. 
In addition, the Landau pole present in the perturbative approach at low 
momentum transfer is screened by a dynamically generated mass. This effect 
is not a particular feature of the present model, but is inherent in the 
standard Gross-Neveu model, too~\cite{la95}. Let us analyze the 
agreement between the large $N_f$ result and the perturbative result 
by expanding the full result in inverse powers of $p^2$, i.e. 
\be 
\Delta _s(p^2) \; = \; 
\frac{ 2 \pi ^2 N_f }{ \ln (p^2 /s_0^2) } \; + \; 
{\cal O} \left( \frac{ M_0^2 }{p^2}, \frac{M^2}{p^2} \right) \; . 
\label{eq:g3} 
\en 
The first term of the right hand side of (\ref{eq:g3}) is precisely 
the result expected from perturbation theory, which is based 
on the existence of free quarks. The confinement order parameter 
$M^2$ only contributes in sub-leading order, i.e. to the operator 
product corrections. The present model provides an example that asymptotic 
freedom is perfectly compatible with the non-liberation of quarks. 
It explains why perturbation theory yields good results in the high 
energy regime, although the perturbative approach is based on free 
quark states.

\bigskip 
{\bf Acknowledgments: } 

I am pleased to acknowledge my collaborator Mannque Rho. It is 
a pleasure to thank Hugo Reinhardt for many inspiring discussions 
as well as for support.

\end{document}